\documentclass[useAMS,usenatbib]{mn2e}
\usepackage{mathrsfs}
\usepackage{graphicx,psfrag}
\usepackage{amsmath,amsfonts,amssymb}

\allowdisplaybreaks

\title[The role of the ergosphere in the Blandford-Znajek process]{The role 
of the ergosphere in the Blandford-Znajek process}
\author[Milton Ruiz, Carlos Palenzuela, Filippo Galeazzi, and Carles Bona]{Milton Ruiz$^{1}$\thanks{E-mail:
milton.ruiz@uib.es}; Carlos Palenzuela$^{2}$\thanks{E-mail:palen@cita.utoronto.ca};
Filippo Galeazzi$^{3}$\thanks{E-mail:filga@aei.mpg.de};
and Carles Bona$^{4}$\thanks{E-mail:cbona@uib.es}\\
$^{1}$Departament de F{\'\i}sica, Universitat de les Illes Balears, Crta. Valldemossa km 7.5, E-07122 Palma, Spain\\
$^{2}$Canadian Institute for Theoretical Astrophysics, Toronto, Ontario M5S 3H8, Canada\\
$^{3}$Max-Planck-Institut f\"ur Gravitationsphysik, Albert-Einstein-Institut,Potsdam-Golm, Germany\\
$^{4}$Departamento de Astronom\'{\i}a y Astrof\'{\i}sica,Universitat de Val\`encia, 46100 Burjassot (Valencia), Spain
}

\begin{document}

\pagerange{\pageref{firstpage}--\pageref{lastpage}} \pubyear{2002}
\maketitle
\label{firstpage}


\begin{abstract}
The Blandford-Znajek process, one of the most promising model for powering the
relativistic jets from black holes, was initially introduced as a mechanism in
which the magnetic fields extract energy from a  rotating black hole.  We study the 
evolution of force-free electromagnetic fields on regular spacetimes with 
an ergosphere, which are generated by rapidly rotating stars. 
Our conclusive results confirm previous works, claiming that the  Blandford-Znajek mechanism is not 
directly related to the horizon of the black hole. We also show that the radiated energy 
depends exponentially on the compactness of the star. 
\end{abstract}

\begin{keywords}
gravitation –- magnetic fields -- relativistic processes -- methods: numerical
\end{keywords}


\section{Introduction}

The Blandford-Znajek (BZ) process is one of the leading models
to explain the launching of powerful relativistic jets emerging from the
supermassive black holes at the center of the galaxies (i.e. Active Galactic
Nuclei), and the more moderated ones coming from stellar mass black 
holes (i.e. microquasars). The main ingredients of this process are 
a central rotating black hole and an accretion disk, which supports a magnetic
field threading the black hole horizon. This magnetic field is twisted by
the spinning black hole, producing an outgoing electromagnetic 
flux which extracts energy and angular momentum from the spacetime.

Although the BZ model was introduced a long time ago (\cite{Blandford02}),
it is only recently that many issues and theoretical discoveries concerning 
this mechanism have been settled. These advances on the understanding 
of the BZ process  have been enabled by numerical simulations. For instance, 
it has been shown that only the magnetic fields lines threading the ergosphere of 
the black hole (i.e. the region near the black hole where negative killing energies 
can exist) rotate due to the frame dragging effect, whether or not they cross the horizon
(\cite{2002astro.ph.11141K,Komissarov:2004ms,2005MNRAS.359..801K,2009JKPS...54.2503K}).
These twisted magnetic fields are carrying the energy of the relativistic jet,
which seems to come from the ergosphere. Moreover, it is now understood the
dependence of the luminosity on the black hole spin magnitude 
(\cite{2010ApJ...711...50T,Palenzuela:2010xn})
and its orientation (\cite{Palenzuela:2010xn}). It has also been shown 
the robustness of the process with respect to different boundary
conditions (\cite{2011CQGra..28m4007P}), and its  resemblance to 
ideal MHD solutions in the limit of high magnetization
(\cite{2004ApJ...611..977M,2005MNRAS.359..801K}).
A generalization of the BZ process to boosted non-spinning black holes has also 
been investigated by~\cite{2010arXiv1012.5661N}, where
the magnetic fields extract the translational kinetic energy from the black holes.
In this case, there is also an extraction of rotational energy through
the original BZ process if the boosted black holes are also spinning.
During the coalescence of binary black hole surrounded by a magnetized
circumbinary disk, this generalized BZ process will produce a dual jet structure during the
inspiral phase which will result into a single BZ jet after the 
merger~(\cite{2010Sci...329..927P}).

The basic effects of the BZ mechanism can be explained by invoking the membrane
paradigm (see~\cite{Thorne82,Thorne86} for details), which endows to the  black 
hole horizon some physical properties like a surface charge density and a resistivity.
The problem is then reduced to a spherical conductor with a relative motion with
respect to asymptotic magnetic field lines via rotation or translation. The magnetic
field is produced by an external source and described by the force-free approximation.
In spite of its simplicity and relative success,
this analogy  does not yet explain the source of the energy, which cannot be assigned
to the horizon due to causality arguments 
(\cite{1989PhRvD..40.3834P,1990ApJ...350..518P}). The membrane  paradigm
implies that the key ingredient of the mechanism is the black hole horizon,
in contrast with the arguments, pointing rather to the ergosphere, presented by Komissarov.
Because the intrinsic marriage of the horizon and the ergosphere on black
hole spacetimes, one could confuse the physical phenomena generated by each
of them. It is therefore desirable to study the effect of each component separately. 

In this paper, we perform a systematic study of the isolated effect of the ergosphere 
in the EM  fields, by considering regular spacetimes produced by rapidly rotating neutron 
stars. By increasing the compactness of the star an ergosphere appears with 
a toroidal topology (see~\cite{Ansorg01} for details). The compact object is immersed in a force-free environment 
produced by an externally sourced magnetic field. We will assume that the force-free EM fields 
are not coupled to the fluid, so their dynamics will be determined only by their evolution equations 
and by the properties of the curved spacetime. Our aim is to analyze the precise role of the ergosphere on
the activation of the BZ mechanism. 

The paper is organized as follows. A detailed description of our model and a
summary of the force-free evolution equations on a curved background is
presented in Section~\ref{section:forandappro}. Some known results 
for stationary and axisymmetric spacetimes are summarized
in Section~\ref{stationary_axisymetric}.
The numerical setup and the initial data is described in
Section~\ref{section:setup}, while that our numerical results are
discussed in Section~\ref{results}. Finally, we summarize our conclusions
in Section~\ref{conclusions}. 
The robustness of our solutions against several sources of error is 
studied in the Appendix.


\section[]{Model of passive force-free environment}
\label{section:forandappro}

We consider the evolution of a magnetized plasma with negligible inertia
on the spacetime produced by a rotating compact star which is assumed to be 
both stationary and axisymmetric. Our approach will
involve the resolution of two different systems of equations. On one hand,
the initial data is obtained by solving the Einstein equations 
coupled to the Hydrodynamic equations. 
We will use an initial data solver
developed by~\cite{Ansorg01} in order to obtain the solution for both the fluid 
and the spacetime geometry. On the other hand, we will evolve the hyperbolic PDE system
for the low-inertia magnetized plasma on this curved background,
which can be described by the force-free approximation of the Maxwell
equations. An important point of our model is that it neglects any coupling 
between the plasma and the fluid of the star. In this way, 
the electromagnetic fields will not interact directly with the fluid,
and its evolution will be determined solely by the force-free equations
in a curved spacetime. In this section, we summarize the formulation used
to describe these systems of equations. In particular, we review in detail
the  Eulerian description of electrodynamics in the force-free approximation. 

\subsection{The 3+1 decomposition}
\label{subsection:3+1formulation}
We consider a spacetime $(M,g_{ab})$ which is foliated by a family of spacelike
hypersurfaces $\Sigma_t$ parametrized by time function $t$.
The induced  metric on these spatial hypersurfaces is denoted by $\gamma_{ij}$.
Coordinates defined on adjacent  hypersurfaces can be related through the lapse
function $\alpha$, that measures the proper time elapsed between both
hypersurfaces, and the shift vector $\beta^i$, that controls how the spatial
coordinates propagate from one hypersurface to the next. 
An observer moving along the normal direction to the hypersurfaces
(Eulerian observer) will have a coordinate speed given by $-\beta^i$,
and will measure a proper time $d\tau = \alpha\,dt$. In terms of these
quantities, it is possible to bring the metric of the spacetime into the form
\begin{eqnarray}
ds^2&=&g_{ab}\,dx^a\,dx^b\nonumber\\
&=&-\alpha^2\,dt^2 + \gamma_{ij}\left(dx^i + \beta^i\,dt\right)\,
\left(dx^j + \beta^j\,dt\right)\,.
\label{eq:line_element}
\end{eqnarray}
Here, and in what follows,  Latin indices from the beginning of the alphabet$
(a, b, c, \cdots)$ denote four-dimensional spacetime quantities, whereas  Latin
indices from the middle of the alphabet $(i, j, k,\cdots)$ are spatial. 
It is also convenient to introduce  the extrinsic curvature $K_{ij}$, which
is associated to the way in which the hypersurfaces are immersed in the spacetime
$(M,g_{ab})$, in the form
\begin{eqnarray}
 K_{ij} &=& -\frac{1}{2 \alpha}(\partial_t - {\cal L}_{\beta})\, \gamma_{ij}\,. 
\label{eq:extrinsic_curvarure}
\end{eqnarray}
Notice that the Eulerian observer is defined independently of the space coordinates.
It can be interpreted as being at rest in the hypersurface $\Sigma_t$. In the
context of spinning stars or black holes, this observer is also called the locally 
non-rotating observer or zero-angular-momentum observer (ZAMO).


\subsection{ 3+1 decomposition of the Maxwell Equations}
The covariant Maxwell equations are given by
\begin{equation}
\nabla_b F^{ab} = 4\,\pi\,I^a\,,
\qquad \nabla_b{} ^*F^{ab} = 0\,,
\label{eq:Maxwell_eqs}
\end{equation}
where $I^b$ is the $4$-current and $F^{ab}$, ${}^*F^{ab}$ are the Maxwell
and the Faraday tensor, respectively.
In order to provide an Eulerian description of the above equations, it is
convenient to introduce the electric and magnetic fields measured by those
observers, namely
\begin{eqnarray}
E^a&=& F^{ab}\,n_b\,,\qquad B^a={}^*F^{ab}\,n_b\,,
\label{eq:magneticfield}
\end{eqnarray}
where $n^a$ is the unit vector normal to the hypersurface $\Sigma_t$.
Notice that, if the electric and magnetic susceptibilities of the medium vanish, as in
vacuum or in a highly ionized plasma, the Faraday tensor becomes the  dual
of the Maxwell tensor. In a similar way, we define the charge density and 
current  as
\begin{eqnarray}
q=-I^a\,n_a\,,\qquad J^a=\perp^a_b\,I^b\,,
\end{eqnarray}
where $\perp^a_b= {\delta^a}_b+n^a\,n_b$ is the projection operator onto the
hypersurface $\Sigma_t$.  Using the previous definitions, the  Maxwell equations 
can be rewritten  in the form
\begin{eqnarray}
(\partial_t - {\cal L}_{\beta})\, E^{i} &=&
\epsilon^{ijk}\,D_j (\alpha \,B_k)+\alpha\,K\, E^i - 4\,\pi\,\alpha\,J^{\,i} \,,
\label{eq:3+1maxwell1} \\
(\partial_t - {\cal L}_{\beta})\, B^{i}&=&
-\epsilon^{ijk}\,D_j (\alpha\,E_k)+\alpha\,K\,B^i \,,
\label{eq:3+1maxwell2} \\
D_iE^i&=&4\,\pi\,q\,,\qquad D_iB^i=0\,.
\label{eq:3+1maxwellconstraints} 
\end{eqnarray}
Here $D_i=\perp^a_i\nabla_a$ is the covariant derivative associated  with the spatial 
metric $\gamma_{ij}$ and $\epsilon^{ijk}$ is the Levi-Civita tensor.

It is useful to introduce, for later convenience, the  vector potential 
$\mathcal{U}_a$ which can be decomposed into
\begin{eqnarray} 
\Phi=-\mathcal{U}_a\,n^a\,,\qquad A_a=\perp^b_a\,\mathcal{U}_b\,.
\label{eq:potentialMaxwell}
\end{eqnarray}
In terms of this vector potential, the Maxwell tensor can be written down as
\begin{eqnarray} 
F_{ab}=-2\,\nabla_{[a}\mathcal{U}_{b]}\,.
\label{eq:Maxpotential}
\end{eqnarray}
On the other hand, the electromagnetic energy-momentum tensor,
\begin{equation}
T_{ab} = \frac{1}{4\pi}\,\left[{F_a}^c\,F_{bc} - \frac{1}{2}\,g_{ab}\,F^{cd}\,F_{cd}\right]\,,
\label{eq:TMaxwell}
\end{equation}
can be decomposed in the form
\begin{eqnarray}
T_{ab}=\mathcal{E}\,n_a\,n_b + 2\,n_{(a}\,S_{b)} + S_{ab}\,,
\end{eqnarray}
where $\mathcal{E}$, $S_a$ and $S_{ab}$ correspond to the local
electromagnetic energy density, the momentum density (Poynting vector) 
and the spatial stress tensor as measured by the Eulerian observer.
Finally, the local conservation of the energy-momentum 
tensor~(\ref{eq:TMaxwell}) is given by
\begin{equation}
\nabla_bT^{ab}=-F^{ab}\,I_b\,.
\end{equation}
The key point about this  discussion is that it has been formulated
in terms of physical quantities measured by the Eulerian observer (ZAMO). 
In order to close the system 
of the Maxwell equations, where a relation between the EM fields and the electric current 
is required, one can use quantities measured by the Eulerian observer in the same way
as in the special relativistic electrodynamics 
(see~\cite{Macdonald:1982,Komissarov:2004ms} for details). 

\subsection{Force-free approximation}
The force-free approximation is valid in magnetized plasmas when the electromagnetic
energy density~$\mathcal{E}$ dominates over matter energy density. It happens, for
instance, in the magnetospheres of neutron stars or black holes, where
the density of the plasma is so extremely low that even moderate magnetic field
stresses will dominate over the fluid pressure gradients. In this limit, the
stress-energy tensor of the plasma therefore satisfies
\begin{equation}
T^{ab}=T_{\textrm{fluid}}^{ab}+ T_{\textrm{EM}}^{ab}\simeq T_{\textrm{EM}}^{ab}\,.
\end{equation}
The local conservation of this stress-energy tensor implies that the Lorentz
force vanishes, $F^{ab}\,I_b\simeq0$ (\cite{Goldreich:1969sb,Blandford02}). 
This expression can be written, in terms of 3+1 quantities, as
\begin{eqnarray}
E^l\,J_l=0\,,\qquad q\,E^l+\epsilon^{ljk}\,J_{j}\,B_k=0\,.
\label{eq:Lorentzforce}
\end{eqnarray}
Taking the scalar and the vector product between the magnetic field $B^i$ and
the spatial projection of the Lorentz force~(\ref{eq:Lorentzforce}), we obtain
\begin{eqnarray}
&&E^lB_l=0\,,\qquad
J^i= \frac{1}{B^2}\,\left(J^i_\parallel+ J^i_\perp\right)\,,
\label{eq:EperJ_Jp}
\end{eqnarray}
where $J^i_\parallel$ and $J^i_\perp$ are the component of the current parallel
and perpendicular  to the magnetic field $B^i$, respectively. These are defined as 
\begin{eqnarray}
J^i_\parallel&=&J^l\,B_l\,B^i\,,\qquad
J^i_\perp=q\,\epsilon^{ijk}\,E_j\,B_k\,.
\end{eqnarray}
The first relation in~(\ref{eq:EperJ_Jp}) implies that the electric and magnetic
field must be perpendicular. The second one defines the current up to the parallel
component $J_\parallel$. Using the Maxwell equations, one can compute $(\partial_t-\mathcal{L}_\beta)(E^l\,B_l)=0$, 
which has to vanish due to~(\ref{eq:Lorentzforce}). This condition imposes a constraint for $J_\parallel$, which can 
be substituted into Eq.~(\ref{eq:EperJ_Jp}) to complete the specification of the
current~(see \cite{Gruzinov:2007qa} for details). We will  use instead an alternative
prescription to enforce the force-free conditions, which has been used successfully 
in previous studies of force-free magnetospheres
(\cite{2006ApJ...648L..51S,Palenzuela:2010xn}).


\section[]{Stationary and axisymmetric spacetimes}
\label{stationary_axisymetric}

In the previous section, we have summarized the Maxwell equations and the force-free
approximation on a generic spacetime. Nevertheless, since we are interested on
stationary and axisymmetric spacetimes, one can consider a set of coordinates adapted to
these symmetries. In these coordinates, the  metric of the spacetime can be brought
into the standard form (\cite{Lewis:1932,Papapetrou66})
\begin{equation}
ds^2=-\alpha^2\,dt^2+ g_{\phi\phi}\,(d\phi-\omega\,dt)^2+ g_{rr}\,dr^2 
+g_{\theta\theta}\,d\theta^2\,,
\label{eq:circularmet}
\end{equation}
where the metric coefficients $\{\alpha,\omega,g_{rr},g_{\theta\theta},
g_{\phi\phi}\}$ depend only on $r$ and $\theta$. 
Note that this metric describes usual astrophysical objects such as neutron
star or black hole spacetimes. In particular, the Kerr metric can be written in the
above form (\cite{Bergamini:2003ch}). As we have mentioned before,  the shift
vector is related to the relative velocity between the Eulerian observer and the
stationary spatial coordinates. One can then interpret $\omega$ as the drag velocity of this
observer with respect to the hypersurface $\Sigma_t$. 

Using the decomposition of the Maxwell tensor in terms of the vector 
potential~(\ref{eq:potentialMaxwell}), the condition of axisymmetry and 
stationarity implies that the electric field is purely poloidal, $E_\phi=0$.
According to~(\ref{eq:EperJ_Jp}), it follows that $E^i$ is perpendicular to the
poloidal components of the magnetic field, so that one can rewrite $E^i$ in the form
\begin{equation}
E^i={\epsilon^i}_{jk}\,B^j\,U^k\,,
\label{eq:EintermsB}
\end{equation}
where $U^k$ is an axial vector given by (\cite{Komissarov:2004ms}) 
\begin{equation}
U^a=\frac{1}{\alpha}\,(\Omega-\omega)\chi^a\,,
\label{eq:EB_speed}
\end{equation}
and  $\chi^a=\partial_\phi$ is the axial killing vector of the spacetime. Notice that,
according to~(\ref{eq:circularmet}),
the shift vector is $\beta^k=-\omega\,\chi^k$. Therefore, one can interpret the
velocity $U^i$ as the velocity of the magnetic field relative to the  Eulerian
observer and $\Omega$ as the angular velocity of the magnetic lines, which
can be written in terms of the Maxwell tensor as (\cite{Blandford02}) 
\begin{equation}
\Omega= \frac{F_{tr}}{F_{t\phi}}=\frac{F_{t\theta}}{F_{\theta\phi}}\,.
\label{eq:OmegaMaxwell}
\end{equation}

It is also useful to calculate the scalar $B^2-E^2$ which, using the electric
field defined by~(\ref{eq:EintermsB}), takes the form 
\begin{eqnarray}
(B^2-E^2)\,\alpha^2&=&{B^2}\,\alpha^2-g_{\phi\phi}\,B_{\mathbf{p}}^2
\,(\Omega-\omega)^2\,,
\label{eq:invariantBE}
\end{eqnarray}
where $B_{\bf{p}}^2= B_r\,B^r+ B_\theta\,B^\theta$ is the magnitude of the 
poloidal component of the magnetic field. This relation implies a change of
sign of this invariant in highly compact rotating spacetimes with large
$g_{\phi\phi}/\alpha^2$ and $\omega$. Notice that, in electrovacuum scenarios,
the Maxwell equations imply that $B^\phi$ vanishes. In this case,
one can also  assume that the magnetic field is generated by distant plasma 
of large inertia, which means that the resulting magnetosphere will reach a steady 
state when $\Omega=0$. This implies that the invariant~(\ref{eq:invariantBE})
becomes
\begin{eqnarray}
(B^2-E^2)\,\alpha^2&=&{B^2_\mathbf{p}}\,\left(\alpha^2-\beta^2\right)\,.
\label{eq:invariantBEelectrova}
\end{eqnarray}
Inside the ergosphere $\alpha^2-\beta^2<0$. Therefore,  the change of the 
sign of the invariant is related, at least in electrovacuum, with the presence 
of an ergosphere.

Finally, it is possible to define conserved quantities associated to the  killing
vectors of the spacetime $\xi^a=\partial_t$ and $\chi^a=\partial_\phi$. On one hand,
the red shifted energy density, corresponding to the killing vector $\xi^a$, is defined as
(see~\cite{Blandford02,Macdonald:1982})
\begin{eqnarray}
\mathcal{E}_\xi=T^{ab}\,\xi_a\,n_b&=&\alpha\,\mathcal{E}+\omega\,S^i\,\chi_i\,,
\end{eqnarray}
with flux of energy given by
\begin{eqnarray}
S^i_\xi=-T^{bc}\,{\perp^i}_b\,\xi_c&=&\alpha\,S^i+\omega\,S^{ij}\,\chi_j\,.
\label{eq:flux-xi}
\end{eqnarray}
On the other hand, the angular momentum density, associated to the killing vector $\chi^i$, is defined
\begin{eqnarray}
\mathcal{E}_\chi=-T^{ab}\,\chi_a\,n_b&=&S^i\,\chi_i\,,
\end{eqnarray}
with a flux of angular momentum given by
\begin{eqnarray}
S^i_\chi=T^{bc}\,{\perp^i}_b\,\chi_c&=&S^{ij}\,\chi_j\,.
\label{eq:fluxredshift}
\end{eqnarray} 
Since $E_\phi$ vanishes, the poloidal flux vector $S^i_\chi$ satisfies
\begin{equation}
S^{\bf{p}}_\chi=-\frac{1}{4\,\pi}\,(B^l\,\chi_l)\,B^{\bf{p}}\,,
\label{eq:energyflux-xi}
\end{equation}
Using this condition and~(\ref{eq:EperJ_Jp}), it is straightforward to show that
\begin{equation}
S^{\bf{p}}_\xi=\Omega\,S^{\bf{p}}_\chi\,.
\label{eq:energyflux-xi}
\end{equation}
Therefore, both the flux of red shifted energy and the flux of angular momentum 
are transported along the poloidal field lines.

According to Eq.~(\ref{eq:flux-xi}), the EM radiated energy crossing a spherical 
surface at a given radius is
\begin{equation}
    \partial_t S=2\,\pi\,\int^\pi_0 \sqrt{-g}\,S^r_\xi\,d\theta\,.
\label{eq:radiated_energy}
\end{equation}
Note that, on a regular spacetime in Lewis-Papapetrou coordinates
(\ref{eq:circularmet}), the radiated energy flux density $S^r_\xi$
is given by
\begin{eqnarray}\label{radiated_energy_LP}
S^r_\xi=-\frac{\Omega}{2\,\pi}\,B^r\,B^\phi\,\alpha^2\,g_{\phi\phi}\,.
\end{eqnarray}
Moreover, on the case of a Kerr spacetime in Lewis-Papapetrou coordinates
(\cite{Bergamini:2003ch}), the above expression becomes
\begin{eqnarray}
S^r_\xi=-\frac{\Omega}{2\,\pi}\,B^r\,B^\phi\,\Delta\,,
\end{eqnarray}
where $\Delta=r^2 + a^2 -2\,M\,r$. Since the Kerr spacetime in these coordinates
is singular at the horizon, it is convenient to
transform to other coordinates that penetrate the horizon smoothly. This is the
case for the Kerr-Schild coordinates, where the energy flux density $S^r_\xi$ can
be written as
\begin{eqnarray}
S^r_\xi&=& \frac{\Omega\,r}{2\,\pi}\,(B^r)^2\,\left(\frac{a}{2\,M\,r}-\,\Omega\right)
\,\sin^2\theta\nonumber\\&&-\frac{\Omega}{4\,\pi}\,\Delta\,B^r\, B^{\phi}\,\sin^2\theta\,,
\label{eq:F_E0}
\end{eqnarray}
At the horizon, where $r=r_H$ and $\Delta=0$, it becomes
\begin{equation}
S^r_\xi|_{r=r_H} = \frac{\Omega\,r_H}{2\,\pi}\, (B^r)^2
\,(\Omega_H - \Omega)\,\sin^2\theta\,,
\label{eq:F_E}
\end{equation}
where $r_H = M + \sqrt{M^2 -a^2}$ is the radius of the horizon and
$\Omega_H \equiv a/(2\,M\,r_H)$ can be interpreted as its rotation
frequency, which is just the rotation velocity of a Eulerian observer at
the apparent horizon.
This result implies that, if $0< \Omega < \Omega_H$ and $B^r \neq 0$, then
there is an outward directed energy flux at the horizon. Therefore, rotational 
energy is being extracted from the black hole due to the magnetic field
lines. The use of Kerr-Schild coordinates allow for direct computations
of the flux at the horizon without any special treatment as 
in~\cite{Blandford02,Macdonald:1982}. However, one message from this
simple calculation is that energy comes out of the event horizon,
which is forbidden at the classical level since the horizon is a null surface.
The problem lies in the fact that the energy flux defined on other
surfaces is not obviously positive definite.


\section{Numerical Set up}
\label{section:setup}


\subsection{Diagnostic Quantities}
\label{section:diagnostic}

To extract physical information, we monitor the rotation frequency of the
magnetic field lines~(\ref{eq:OmegaMaxwell}) which is constant along magnetic
fields lines on axisymmetric and stationary
solutions (\cite{Blandford02}), and the Newman-Penrose electromagnetic scalars
$\{ \Phi_0, \Phi_2 \}$, which are computed by contracting the Maxwell tensor
with a suitable null tetrad~(see~{\it{e.g.}}~\cite{Teukolsky:1973ha}),
\begin{equation}
\Phi_0 \equiv-F^{ab} m_{a} l_{b}\,, \qquad
\Phi_2 \equiv F^{ab} \overline m_{a} n_{b} \,.
\end{equation}

The total energy flux (luminosity) of electromagnetic waves,
which accounts for the energy carried off by outgoing waves to infinity, is
\begin{equation}
L_{EM} = \lim_{r \rightarrow \infty}  \int  
r^2 \left( |\Phi_2 - \Phi^B_2|^2 - |\Phi_0 - \Phi^B_0|^2 \right) d\Omega\,,
\label{EM_luminosity_general}
\end{equation}
where $\Phi^B_2$ and $\Phi^B_0$ are the background scalars produced by
the steady part of the solution, which vanish only at far distances
from the electromagnetic sources. 
However, since we are considering for simplicity that the magnetic field
is produced by a very distant external source, there will be a non-zero
contribution to these background scalars induced by the asymptotically
uniform magnetic field configuration. An isolated system with no incoming
radiation satisfies $\Phi_0 = \Phi^B_0$. Moreover, far from the
star is valid the assumption that the background is approximately
the same for the incoming and outgoing waves, so that $\Phi^B_2 \approx \Phi^B_0$.
Combining these relations with the general form given
by~(\ref{EM_luminosity_general}), it is obtained the simplified formula
\begin{equation}\label{EM_luminosity_CP}
L_{_{\rm EM}} = \lim_{r \rightarrow \infty}  \int  
r^2 |\Phi_2 - \Phi_0|^2  d\Omega\,,
\end{equation}
which has been used previously in several works, reproducing 
successfully the expected analytical relations 
(\cite{Palenzuela:2009yr,Palenzuela:2010xn,2010arXiv1012.5661N}). Notice that these
expressions are equivalent to the radiated energy (\ref{eq:radiated_energy})
evaluated at spatial infinity.


\subsection{Numerical methods}
\label{subsection:methods}
We will use a  finite difference scheme on a regular Cartesian grid to
solve numerically the hyperbolic PDE system.
To ensure sufficient resolution in an efficient manner we employ 
Adaptive Mesh Refinement (AMR) via the {\sc had}
computational infrastructure, that provides distributed,
Berger-Oliger style AMR (\cite{hadcode,Liebling:2002qp}) with full 
sub-cycling in time, together with an improved treatment of artificial
boundaries as it has been presented by~\cite{Lehner:2005vc}. For these
simulations, the refinement regions are fixed initially and not changed
during the evolution (i.e. Fixed Mesh Refinement).

The spatial discretization is performed by using a fourth order accurate scheme 
satisfying the Summation By Parts rule. The time evolution is performed through 
the method of lines using a third order accurate Runge-Kutta integration scheme with
a Courant parameter of $\lambda= 0.25$ such that 
$\Delta t=0.25\,\Delta x$ holds in each refinement level.

Our numerical domain consists of a cubical region defined by the intervals
$x^i \in [−32\,M, 32\,M]$ with $61$ points in the coarsest grid.
We employ a fixed
mesh refinement configuration with 6 levels of  refinement, each one
covering half of the domain of the parent coarser level. 
The coarsest resolution employed is $\Delta x=1.07$ while the finest one
is $\Delta x = 0.017$. The radius of the different stars in these units
are described in Table~\ref{tab:ergoID}.
We have adopted maximally dissipative boundary conditions in our simulations,
by setting to zero the (time derivative) of the electrovacuum incoming modes
(\cite{2011CQGra..28m4007P}).

\subsection{Initial data}
\label{subsection:id}

The initial data for the spacetime geometry and the fluid variables
produced by rotating stars is obtained by solving the Einstein and
the Hydrodynamic equations with the assumptions of stationarity and
axisymmetry. 
The rotating star solutions have been constructed using the code developed 
by~\cite{Ansorg01} based on a multi-domain spectral-method for representing
the metric functions. The use of a spectral code has shown to be necessary 
to achieve high accuracy in the case of a stiff equation of state (e.g. for constant
total mass-energy density~\citep{Bonazzola1974}). We consider equilibrium solutions 
for a rigidly rotating star with an equation of state for homogeneous matter with constant total 
mass-energy density, $\mu=\text{const}$.
For the calculation we use two different line elements to describe the exterior and
the interior of the star.
The Lewis-Papapetrou line element~(\ref{eq:circularmet}) that covers the exterior 
has the form
\begin{eqnarray}\label{LP_element_line}
ds^2 &=& - e^{2\,\nu}\,dt^2 + W^2\,e^{-2\,\nu}\,(\omega\,dt-d\phi)^2 \nonumber\\ 
&&+e^{2\,\alpha}\,(d\rho^2+ d\xi^2)
\end{eqnarray}
The advantage of this line element is that allows the metric potential $\nu$ 
to remain real inside the ergosphere. For the interior of the star, in 
the comoving frame of coordinate, the metric can be expressed as
\begin{eqnarray}\label{SCF_element_line}
ds^2 &=&- e^{2U}\,dt^2 +e^{-2U} \left[ e^{2\,k}\,(d\rho^2+d\xi^2) \right.
\nonumber\\&+&\left. (W^2+\eta)\,d\phi^2\right]\,.
\end{eqnarray}
The potential $U$ can be expressed in terms of the lapse function $\alpha$, while $\eta$ is the 
so-called gravitomagnetic potential associated with the shift vector (see ~\cite{AnsorgBook08} 
for a detailed description).
Given the particular equation of state and using the conservation of the 
energy-momentum tensor for the fluid,  we obtain inside the star
\begin{equation}
e^{U}\,\text{exp}\,
\left[\int_0^{p} \frac{dp}{\mu+p}\right]=e^{V_0}
=\text{const}\,.
\end{equation}
Isobaric surfaces inside the star correspond to a constant value of $V_0$. At the surface,
where pressure goes to zero, it is possible to compute the redshift of a photon emitted with zero 
angular momentum via
\begin{eqnarray}
z= e^{-V_0} -1\,.
\end{eqnarray}
By changing the parameter $V_0$, the solution becomes more compact and may
contain an ergosphere. We have constructed several rotating stars, with
different value of $V_0$ and rotation frequency $\Omega$. We kept the ratio
between polar and equatorial radius constant.  
For all the models the value of the dimensionless spin parameter is roughly
constant, $a=J/M^2 \approx 0.9$. The mass, radius and
other parameters of the solutions are displayed in Table~\ref{tab:ergoID},
where all the solutions listed in the table contain an ergosphere. Our most
compact star is close to the limit of maximum compactness
$M/R < 4/9 \approx 0.44$ for this family of solutions.

Stationary and asymptotically flat configurations with an ergosphere but without an 
horizon have been proved to be unstable or marginally unstable under to scalar and
electromagnetic perturbations (\cite{Friedman1978}). For slowly rotating  
relativistic stars, the time scale of the instability is shown to be longer than the Hubble 
time (see e.g.~\cite{Comins1978}). It has been shown by~\cite{Cardoso2007} that, for the 
extreme case of compactness $M/R > 0.5$ and angular momentum $J  >0.4M^2$, 
the instability timescales reaches 0.1 seconds for an object with mass of $1M$.
In our simulations, however, both the spacetime and the fluid are stationary and therefore
this instability cannot be active.

The initial data for the black hole is analytical. In the case of spinning
black holes, we will use the Kerr-Schild coordinates 
(\cite{Kerr1963}, see {\it e.g.}~\cite{KraHer80} for more details),
\begin{eqnarray}
g_{ab}= \eta_{ab} + 2\,H\,l_a\,l_b   \,,
\end{eqnarray}
where $\eta_{ab}$ is the Minkowsky metric and the scalar function $H$ and the null
vector $l^b$ are defined by
\begin{eqnarray}
H&=&\frac{r\,M}{r^2 + a^2\,z^2/r^2}\,,\\
l_b&=&\left(1,\frac{r\,x+a\,y}{r^2+a^2},\frac{r\,y-a\,x}{r^2+a^2},\frac{z}{r}\right)\,.
\end{eqnarray}
The compact object, either a neutron star or a black hole, is immersed
in the external magnetic field produced by a distant current loop.
This magnetic field is nearly constant initially near the compact object.
In addition, it is chosen to be aligned with the spin of the compact
object, which is initially oriented along the z-axis. Therefore,
the EM fields are initially set to $B^i= B_o\,\hat{z}$ and $E^i=0$
throughout all the domain. The field strength $B_o$ is irrelevant, since
we are assuming that the force-free fields behaves like test fields
(i.e. they do not modify the curvature of the spacetime),
and it has been set to $B_o = 0.01$.
Since we are not considering any coupling between the fluid and the
force-free EM fields, the dynamics of the latter will be only
influenced by the regular spacetime both {\it inside and outside}
the star. For all the effects, there will be no direct interaction
between the EM fields and the fluid.

\begin{table*}
 \begin{minipage}{140mm}
\caption{\label{tab:ergoID} $M_b$ baryonic mass, $M_{ADM}$
gravitational mass, $J$ angular momentum, $\Omega$ angular 
velocity of the fluid, $r_e$ equatorial radius, $r_c$ circumferential radius in Schwarzschild
coordinates, $w_c$ metric potential and $V_0$ 
parameter. All the configurations contain an ergosphere.The ratio between polar 
and equatorial radius, $r_p/r_e$ , is equal to $0.49$ for all the configurations.
All numerical values are in units with $G = c = 1$.}
 \centering
\begin{tabular}{|c|c|c|c|c|c|c|c|c|}
  \hline
   $V_0$ &  $M_b$    &   $M_{ADM}$  & $J$   & $\Omega$ & $r_e$ & $r_c$  & $w_c$ \\
  \hline
  \hline
  $-1.00$   & 0.1814 &  0.1395  & $ 0.0177 $  &  $1.5685$ & $ 0.1902$ & $0.3553$& $1.360$ \\
  $-1.05$	 & 0.1885 &  0.1440  & $ 0.1891 $  &  $1.5844$ & $ 0.1857$ & $0.3576$& $1.395$ \\
  $-1.10$   & 0.1998 &  0.1509  & $ 0.0209 $  &  $1.6103$ & $ 0.1782$ & $0.3610$& $1.450$ \\
  $-1.15$   & 0.2018 &  0.1521  & $ 0.0212 $  &  $1.6151$ & $ 0.1768$ & $0.3616$& $1.460$ \\
  $-1.20$   & 0.2080 &  0.1576  & $ 0.0223 $  &  $1.6298$ & $ 0.1725$ & $0.3661$& $1.490$ \\
  \hline
\end{tabular}
\end{minipage}
\end{table*}


\section[]{Results}
\label{results}

In this section we will describe the dynamics of the force-free
fields evolving in the stationary spacetimes produced by very compact
rotating objects with dimensionless spin parameter $a \approx 0.9$.
We will concentrate  on the cases with the presence of an ergosphere
in the spacetime (see the Appendix for a discussion
on the cases without ergosphere). We will also analyze the EM power
(if any) emitted by the BZ process in these spacetimes, as well as the
features of the EM fields after they have relaxed to the stationary
solution.

\begin{figure*}
    \includegraphics[width=0.35\textwidth]{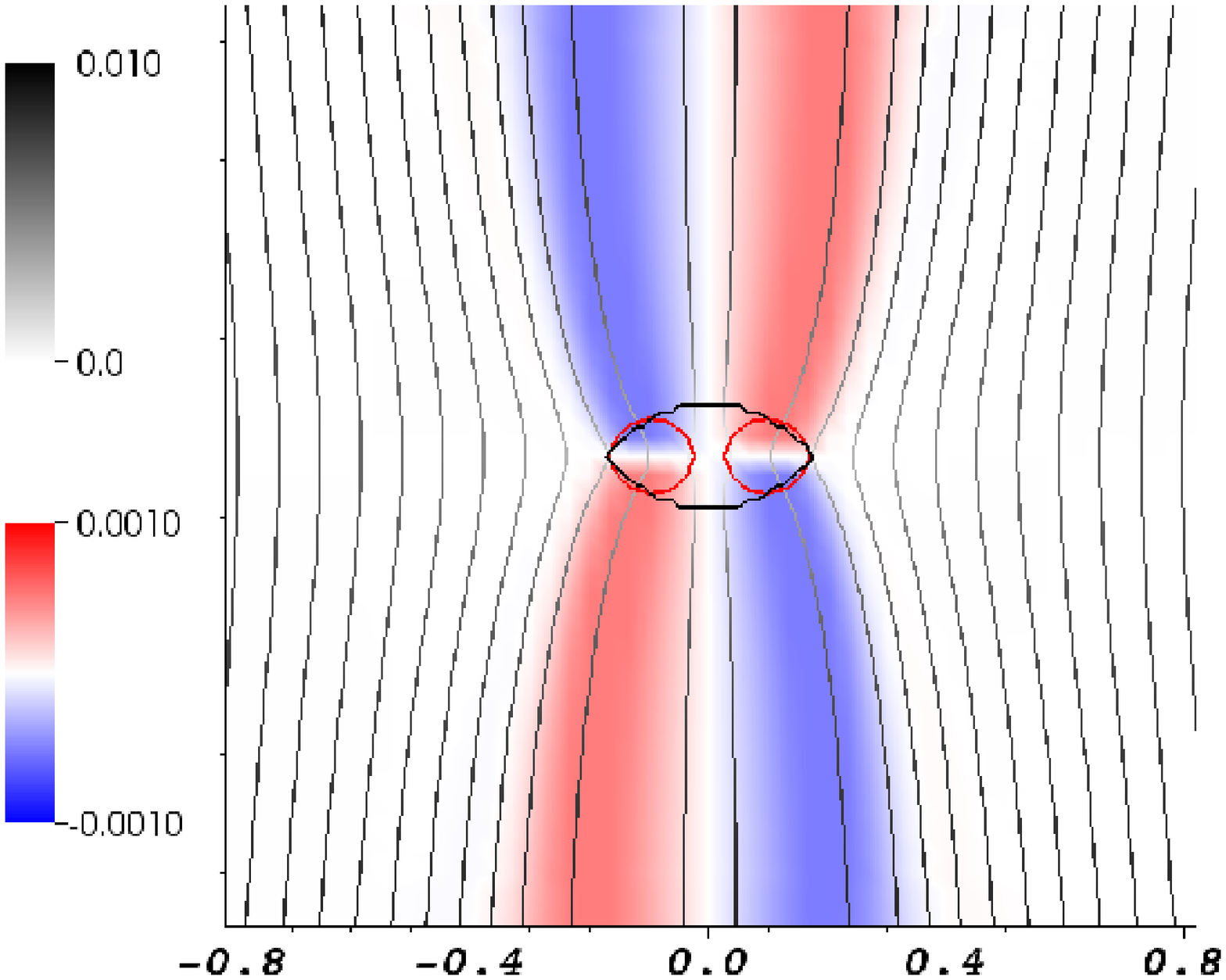}\hspace{0.5cm}
    \includegraphics[width=0.35\textwidth]{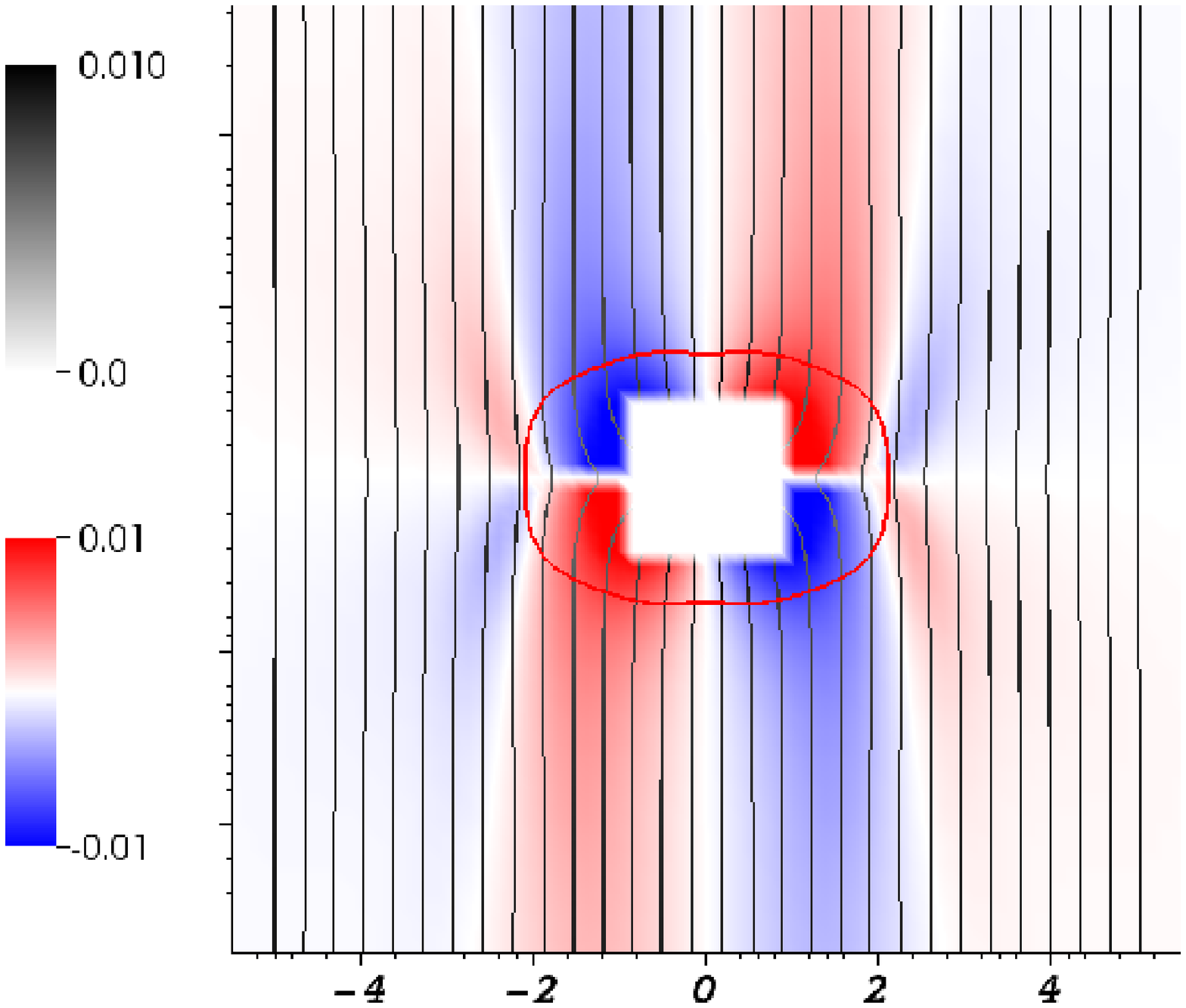}
    \caption{ \label{fig:ergo08-ByzF1_field} 
    {\em Rotating star and black hole}. Magnetic field lines component in
     the plane $x=0$ after the quasi-stationary state is reached, 
     corresponding to the spacetime
     with an ergosphere (left panel) and to the black hole (right panel). The vertical lines
     indicate the poloidal component, while that the blue-red colors indicates
     the strength of the component normal to the plane. 
     The structure of these two components of the magnetic field for both cases, 
     star and black hole, are quite similar each other.
     The surface of the  star is plotted as a black ellipsoid, while that the
     ergosphere is plotted in red.}
\end{figure*}

\begin{figure*}
    \includegraphics[width=0.33\textwidth]{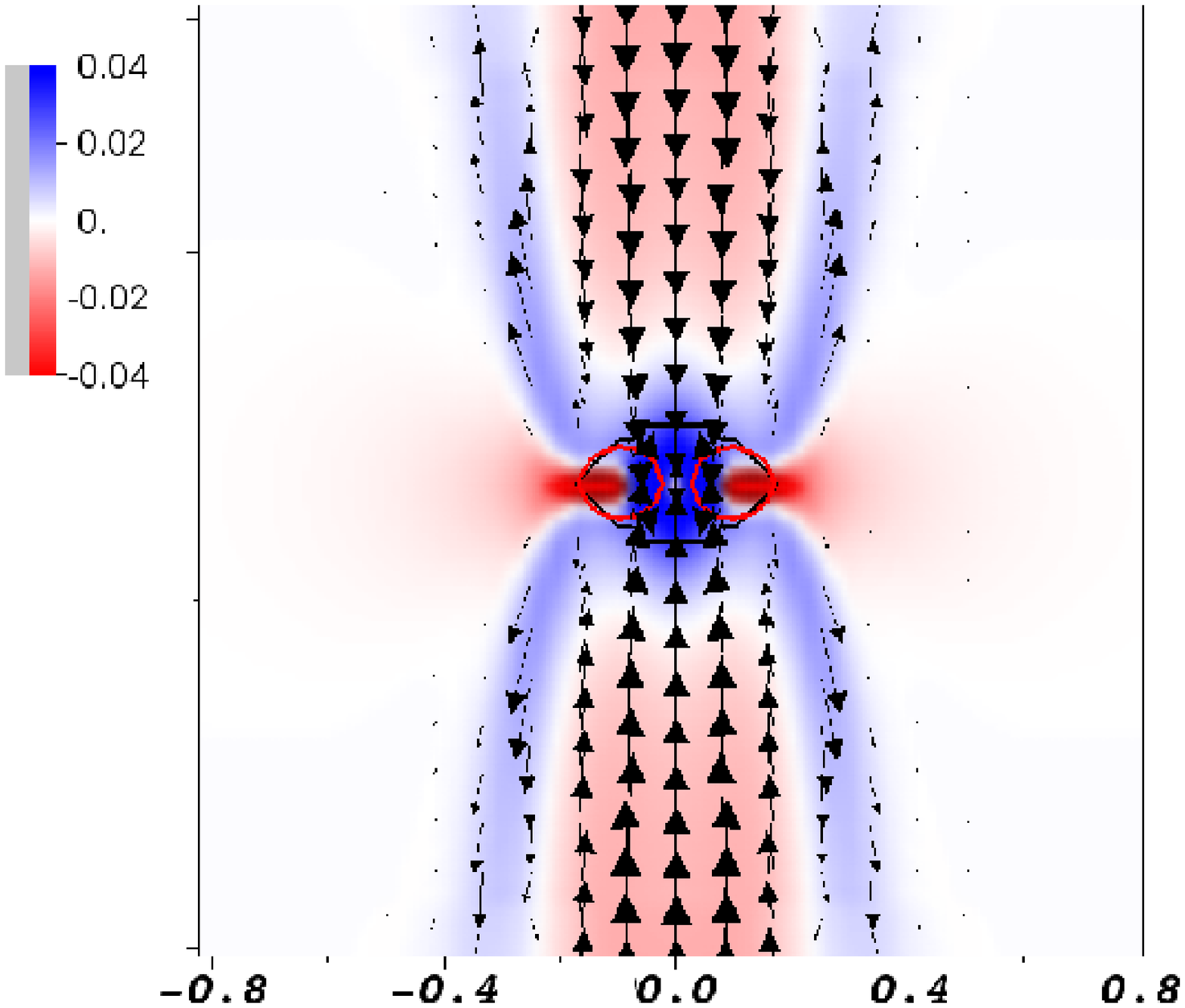}\hspace{0.5cm}
    \includegraphics[width=0.34\textwidth]{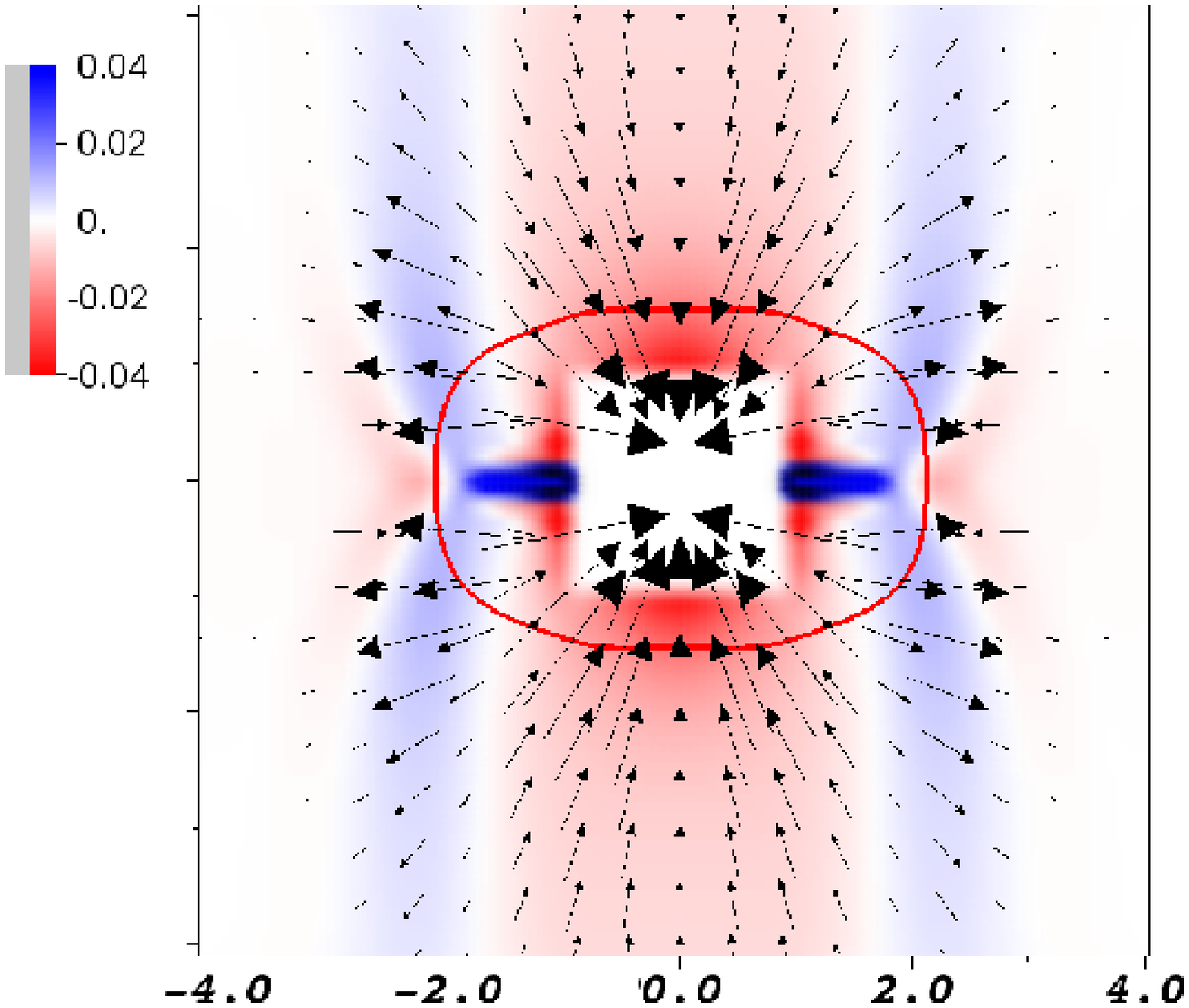}
    \caption{ \label{fig:erg0125-Jstar_field} 
    {\em Rotating stars and black hole}. Induced charge density (in red-blue
     colors) and the poloidal currents (in vectors) on the plane $x=0$ at the
     quasi-stationary state, corresponding to the same cases than
     figure~\ref{fig:ergo08-ByzF1_field}. The surface of the star is plotted
     in black and the ergosphere in red lines.
     }
\end{figure*}

All our simulations display an initial transient, in which the magnetic
field is dragged and twisted around the spinning spacetime and induces
a poloidal electric field. At late times the EM field relaxes to a
stationary state, which is displayed in figure~\ref{fig:ergo08-ByzF1_field}
for a representative case of regular spacetime with an ergosphere.
This case corresponds to $V_0=-1.20$ (see Table~\ref{tab:ergoID}). 
For comparison purposes, we have additionally included the results for the black hole 
case, a spacetime with an ergosphere but also with an horizon which hides a singularity.
Further information on the structure of the solutions can be inferred from the
currents and charge density of these two cases, as displayed in
fig.~\ref{fig:erg0125-Jstar_field}. 

In the case of regular spacetimes with an ergosphere, the magnetic flux near the star
is  initially expelled, presenting large damped oscillations that relaxes 
after few periods.
During this relaxation, which seems to be more
relevant as the compactness of the star increases, there is an important
isotropic emission of energy. When the stationary state is reached, all the magnetic 
fields from the region occupied by the star are twisted in the same direction than 
its angular momentum (in the $z>0$ domain). The currents, in this case, are composed of
an outflow external cylinder and an inflow inner one.
There is also a current sheet where
$B^2 \lesssim E^2$ in the intersection of the ergosphere  with the equatorial plane,
similar to the one that appears in the black hole case 
(\cite{Komissarov:2004ms,Palenzuela:2010xn}).

The black hole simulation, on the other hand, relaxes to the stationary state in a shorter
timescale than the above case. The final state resembles the solution corresponding
to the regular spacetime with an ergosphere, displaying an analogous structure of
magnetic fields, currents and charge densities. This clearly indicates that the
BZ mechanism acting on the spacetimes with an ergosphere is basically the same
than in the black hole case.

\begin{figure*}
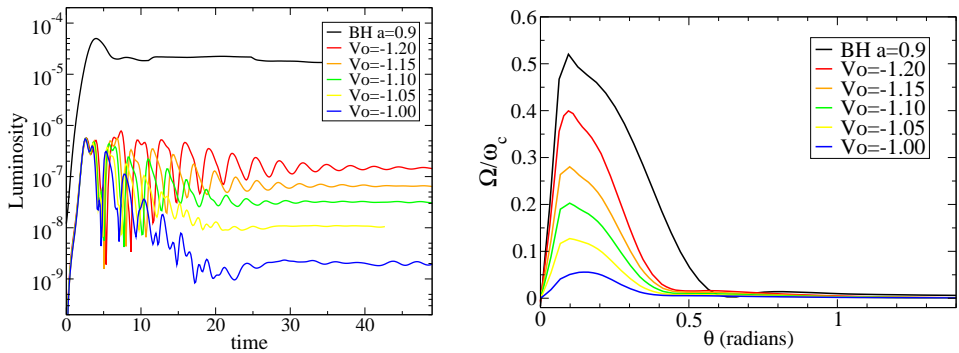

    \includegraphics[width=0.32\textwidth]{luminosity.eps}\hspace{0.5cm}
    \includegraphics[width=0.36\textwidth]{omega.eps}
    \caption{ \label{fig:luminosity} Left panel displays the EM luminosity obtained in the
rotating spacetimes with $a = J/M^2 \approx 0.9$. The luminosity
increases monotonically with the compactness.
Right panel displays the angular velocity of the
magnetic field $\Omega$, computed at $r\approx 5 r_e$ and normalized
with respect to its maximum value inside the star
(see Table~\ref{tab:ergoID}).} 
\end{figure*}

The poloidal structure of the magnetic fields is almost identical in all
the simulations, showing that the magnetic flux threading the spacetime
occupied by the compact object is basically the same.
The luminosity, evaluated in a sphere located at $R \approx 10\,r_e$ for
the stars, and conveniently rescaled for the black hole, is displayed in
the left panel of fig.~\ref{fig:luminosity} for all the simulations. The luminosity increases
very fast as the compactness of the star increases, although it does
not reach the high values of the black hole.  


\begin{figure*}
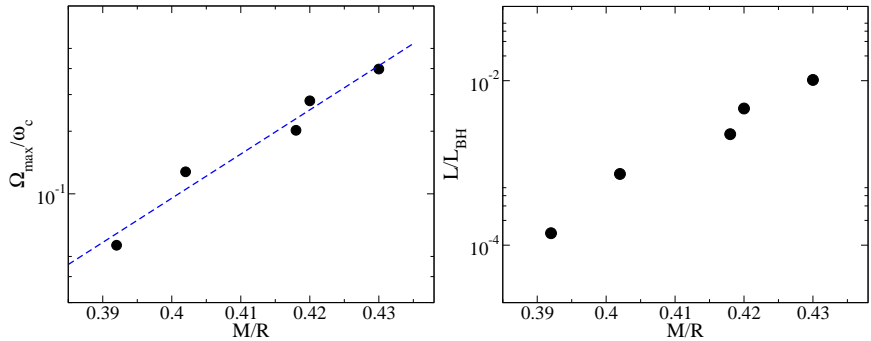

    \includegraphics[width=0.32\textwidth]{Omega_compact.eps}
    \includegraphics[width=0.32\textwidth]{L_compact.eps}
    \caption{ \label{fig:compactness} (Left panel) The maximum of the (normalized)
angular velocity of the magnetic field $\Omega/\omega_c$,
as a function of the compactness. 
A fit with the numerical results shows an exponential dependence
$A\, e^{\lambda\, M/R}\,$ with $A \approx
3 \times 10^{-10}$ and $\lambda \approx 50$. (Right panel) The EM luminosity calculated
in the regular spacetimes, normalized with respect to the black hole one.
Roughly, it seems to depend also exponentially on the compactness.} 
\end{figure*}

This smoothness is also found in the angular velocity of the magnetic field,
displayed in the right panel of fig.~\ref{fig:luminosity}, where $\Omega$ has been normalized with
respect to the central maximum value $\omega_c$ for the stars, and with
respect to $\Omega_H$ for the black hole. As it was mentioned before, the
angular velocity $\Omega$ is confined to a small cylinder, showing that the 
jet is collimated to the region occupied
by the compact object. The fast growth of the maximum of this  quantity
as a function of the compactness of the star can be fitted accurately in
this regime to  an exponential function, as it is shown in the left panel of 
figure~\ref{fig:compactness}. 
The luminosity for the different cases can also be represented as a 
function of the compactness, showing roughly also an exponential 
dependence in the right panel of figure~\ref{fig:compactness}.

From our numerical results we have found the following scaling relations
for the angular velocity $\Omega$,
\begin{equation}
\Omega/\omega_c \approx A\, e^{\lambda\, M/R}\,,
\label{approx_omega}
\end{equation}
and for the ratio of poloidal and toroidal components of the magnetic field
\begin{equation}
B^{\phi} \approx - f\, \Omega B^r\,,
\label{approx_bphi}
\end{equation}
with $f \approx 1/5$ 
for the spacetime with an ergosphere.  Notice that these estimates contain large sources of error,
since they both neglects  the details of the spacetime geometry and the azimuthal
dependence of these quantities. Nevertheless, they can be used to study the
behavior of the solution in different limits and to obtain the right order of
magnitude of the luminosity. By using the line element of our initial data
(\ref{LP_element_line}), the energy flux density (\ref{radiated_energy_LP}) reduces to
\begin{eqnarray}
 S^r_\xi &=& -\frac{\Omega}{2\,\pi}\,B^r\,B^\phi\,W^2\, 
 \approx \frac{f \Omega^2}{2\,\pi}\,(B^r)^2\,W^2\, \nonumber \\
 &\approx& \frac{A f \omega_c^2}{2\,\pi}\,(B^r)^2\,W^2\, e^{2\, \lambda\, M/R}\,, 
\label{approx_energy_flux}
\end{eqnarray}
which is a positive definite quantity, and  where we have used subsequently the above 
approximations. Notice that the scaling
is similar to the BZ power in Eq.~(\ref{eq:F_E}), except by the exponential dependence
on the compactness, and is consistent with the results displayed at the right panel of
fig.~\ref{fig:compactness}.


\section[]{Concluding Remarks}
\label{conclusions}

We have studied the evolution of EM fields on rotating and  highly compact regular spacetimes
with an ergosphere. Our results show that if a ergosphere is present 
the structure of EM fields and currents are  similar to the black hole ones. This implies
that the same mechanism operates in both spacetimes, independently
of the presence/absence of an horizon. Notice that these results are in agreement with
the fact that the BZ process is not an effect caused by the horizon, as it was
pointed out by~\cite{Komissarov:2004ms,2005MNRAS.359..801K} in the context of
black holes.

In the case of a realistic rotating star, the fluid will be coupled with the EM fields and, 
therefore, they are forced to rotate following the fluid. So,  the extraction process 
can be  described by the simple model of the Faraday disk, which is even more efficient than 
the BZ process. In this case, the twist on the magnetic field will not be dominated by the 
spacetime but by the fluid rotation.  In general, the strong dependence of $\Omega$ on the 
compactness will imply a low energy extraction on regular spacetimes even if matter is not 
present.  

An example of regular and rotating
spacetime may be produced by an orbiting binary black hole system, 
as it was considered by~\cite{2011PhRvD..83f4001L}.
In this work it was suggested that the rotation of the spacetime inside  the binary system,
induced by the orbital motion of the black holes, would allow for two different channels of
energy extraction; through the generalized BZ in boosted (and maybe rotating) black holes,
which will produce a dual jet structure (\cite{2010Sci...329..927P}), and through the BZ
process on regular spacetimes inside the binary system. However, the lack of an ergosphere
in the central region will probably prevent the activation of the BZ process
(see the Appendix for a discussion on the case of spacetimes without ergosphere).
Even if the BZ process takes place, and assuming that the luminosity still follows
the exponential dependence shown earlier, the low compactness
of the spacetime in this central region will induce a faint jet that will be overshined by the
BZ process taping kinetic energy (either translational or rotational) directly from the black
holes.

Summarizing, we have found that the BZ process is also present on regular spacetimes 
with an ergosphere. Our conclusive results implies that
we have to reconsider the membrane paradigm as a tool to explain the BZ mechanism
which also seems to be able to extract energy from rotating regular spacetimes
with ergospheres and boosted non-rotating black holes.

\section*{Acknowledgments}
We are in debt with L. Lehner and T. Garrett for stimulating discussions
about the origin of the BZ process which initiated this project.
We are also grateful to Marcus Ansorg and his collaborators 
for giving us the access to his spectral code and for the useful 
discussions on spectral method during his period at AEI.
It is a pleasure to thank L. Lehner, S. Liebling, S. Komissarov,
C. Thompson, as well as to our long term collaborators M. Anderson,
E. Hirschmann and D. Neilsen for useful discussions
and comments on this paper.
We acknowledge support from Spanish Ministry of Science and Innovation
under grants CSD2007-00042, CSD2009-00064 and FPA2010-16495, and 
Govern de les Illes Balears. Computations were performed in Scinet
and Mare Nostrum (funded by BSC Grant No. FI-2011-3-0017).

\bibliographystyle{mn2e}             
\bibliography{refs/references}{}     

\appendix

\section[]{robustness of the results}
\label{sec:appendix}

\begin{figure*}
  \includegraphics[width=0.42\textwidth]{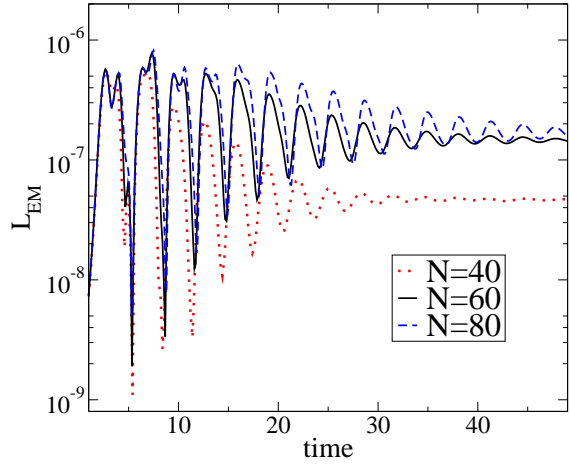}
  \caption{ \label{fig:robustness} 
    The EM luminosity for the cases $V_0=-1.2$ (with ergosphere) for three different spatial numerical
    resolution with $N=\{40,60,80\}$ points in the coarsest grid. The luminosity converges with
    the expected fourth order convergence}
\end{figure*}
We have checked the robustness of our solutions against several sources
of error. One of the possible problems may come from the way in which the 
analysis quantities are evaluated. In particular, the surface where the
luminosity is computed may be located too close to the source, where the
spacetime is still far from being flat, producing an error in that measure.
We have compared the luminosity computed in two surfaces located
and $5\,r_e$ and $10\,r_e$ for the case with $V_0=-1.2$,
obtaining a difference smaller than $5\%$.
Another potential problem may come from the influence of the boundary
conditions, which may produce unphysical reflections which may affect,
after a light crossing time, the dynamics of the system. In our simulations
this is not a problem since the solution relaxes to the stationary state
before a light crossing time, and it remains unaffected afterwards.

Probably the most important source of inaccuracies comes from
numerical discretization errors. We have compared three different spatial
resolutions, corresponding to $N=\{40,60,80\}$ points in the coarsest grid.
Our comparisons are summarized in fig.~\ref{fig:robustness}, where
we have restricted our analysis to a representative case
corresponding to $V_0=-1.2$. Notice that, in this case, there is an
ergosphere, and consequently, a current sheet on the equatorial plane
which is difficult to represent on a discretized grid. Nevertheless,
the luminosity display the expected fourth order convergence to a well
defined solution. 

\begin{figure*}
  \includegraphics[width=0.34\textwidth]{conv_noergo.eps} \hspace{0.5cm}
  \includegraphics[width=0.36\textwidth]{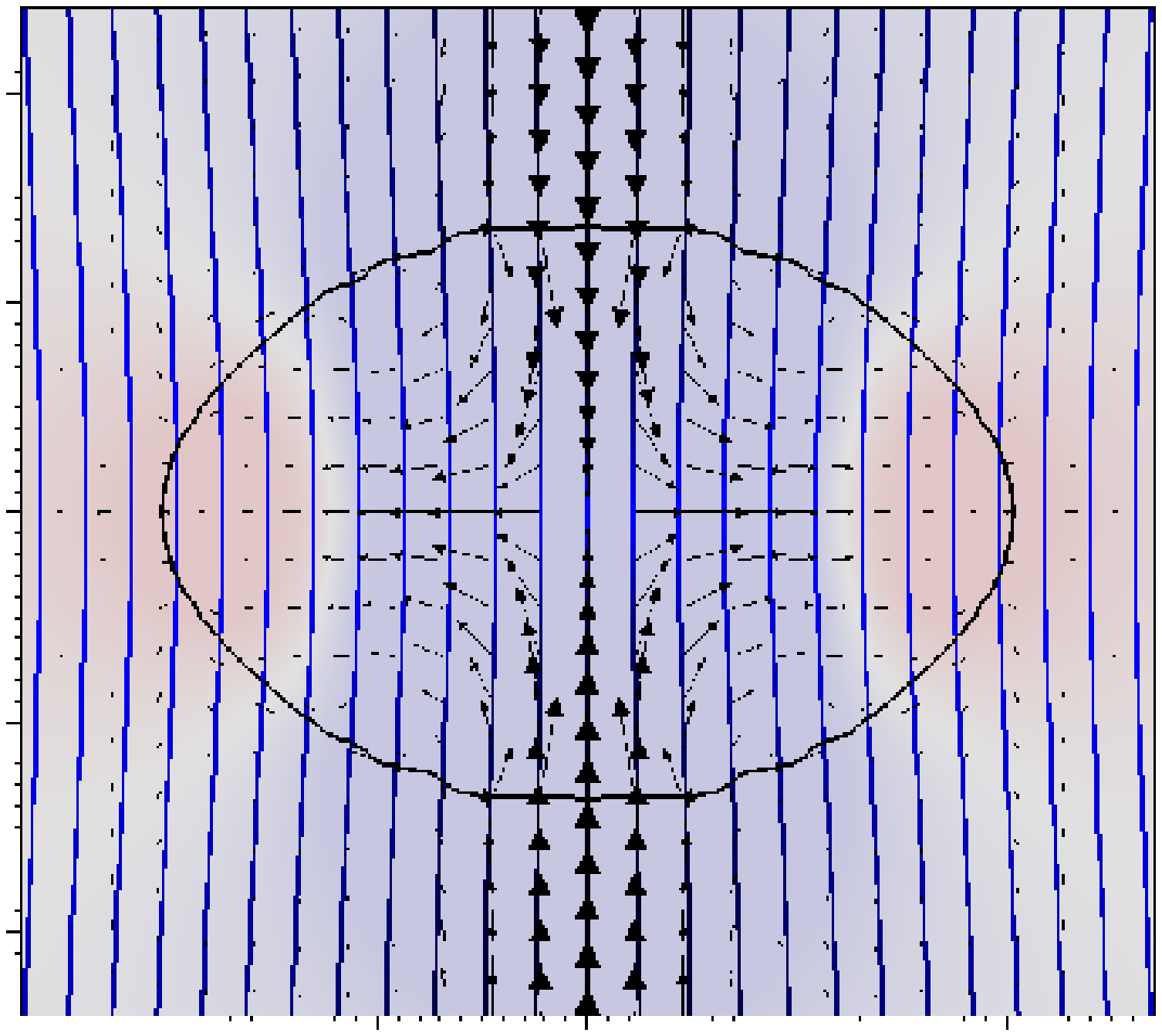} 
  \caption{ \label{fig:noergo} 
    (Left panel) EM luminosity for the same three different resolutions
    as in fig.~\ref{fig:robustness}. In this case the solutions 
    do not display any convergence.
    (Right panel) Poloidal magnetic field lines (in blue) and
    poloidal currents vectors (in black) in
    the plane $x=0$ after the quasi-stationary state is reached, 
    corresponding to the spacetime without an ergosphere. The blue-red
    colors indicates
    the charge density, while that the black ellipsoid represents the
    surface of the star. Although the poloidal component of these fields has
    to be parallel in a force-free stationary and axisymmetric solution,
    it is clearly not satisfied in the interior of the star.}
\end{figure*}

We have also tried to study the relaxed solutions of spacetimes without
an ergosphere. Although the luminosity reaches a quasi-stationary value, 
it changes dramatically with resolution and it does not seem to converge
to a unique solution. This lack of convergence is shown in the left panel
of fig.~\ref{fig:noergo} for a representative case without ergosphere
corresponding to $V_0=-0.97$. In these cases there is a violation of the 
force-free condition, as it can be seen in the right panel of figure ~\ref{fig:noergo}.
The poloidal currents inside the 
star are not parallel to the poloidal magnetic field, a consequence of
a non-vanishing toroidal electric field. This component has to vanish
in stationary and axisymmetric spacetimes. Our guess is that there is
no force-free solution with these boundary conditions unless a current
sheet appears, and so the numerical evolution relaxes only to an
approximated solution which depends strongly on the resolution. More
work is needed to elucidate this issue, maybe considering the full MHD
problem instead of the force-free limit.

\bsp
\label{lastpage}
\end{document}